\documentclass[]{aa} %
\usepackage{graphicx}
\usepackage{txfonts}
\usepackage{verbatim}
\usepackage{bm}

\begin{document}


\title{A horn-coupled millimeter-wave on-chip spectrometer based on Lumped Element Kinetic Inductance Detectors}

\author{U. Chowdhury \inst{1,2}
	\and
	F.~Levy-Bertrand \inst{1,2}
	\and
	M.~Calvo \inst{1,2}
	\and
	J.~Goupy \inst{1,2,3}
	\and
	A.~Monfardini \inst{1,2}
	}

\institute{Univ. Grenoble Alpes, CNRS, Grenoble INP, Institut N\'eel, 38000 Grenoble, France
	\and
	Groupement d'Int\'er\^et Scientifique KID, 38000 Grenoble and 38400 Saint Martin d'H\`eres, France
	\and
	Institut de RadioAstronomie Millim\'etrique (IRAM), 38400 Saint Martin d'H\`eres, France
	}

 
  \abstract
   {
\emph{Context.} Millimetre-wave astronomy is an important tool for both general astrophysics studies and cosmology. A large number of unidentified sources are being detected by the large field-of-view continuum instruments operating on large telescopes.
   
\emph{Aims.} New smart focal planes are needed to bridge the gap between large bandwidth continuum instruments operating on single dish telescopes and the high spectral and angular resolution interferometers (e.g. ALMA in Chile, NOEMA in France). The aim is to perform low-medium spectral resolution observations and select a lower number of potentially interesting sources, i.e. high-redshift galaxies, for further follow-up.   
   
\emph{Methods.} We have designed, fabricated and tested an innovative on-chip spectrometer sensitive in the 85-110~GHz range. It contains sixteen channels selecting a frequency band of about 0.2 GHz each. A conical horn antenna coupled to a slot in the ground plane collects the radiation and guides it to a mm-wave microstrip transmission line placed on the other side of the mono-crystalline substrate. The mm-wave line is coupled to a filter-bank. Each filter is capacitively coupled to a Lumped Element Kinetic Inductance Detector (LEKID). The microstrip configuration allows to benefit from the high quality, i.e. low losses, mono-crystalline substrate, and at the same time prevents direct, i.e. un-filtered, LEKID illumination.
   
\emph{Results.} The prototype spectrometer exhibit a spectral resolution $R = \lambda / \Delta\lambda \approx 300$. The optical noise equivalent power is in the low $10^{-16}$~W.Hz$^{-1/2}$ range for an incoming power of about 0.2~pW per channel. The device is polarisation-sensitive, with a cross-polarisation lower than 1\% for the best channels.
   }



 \titlerunning{Millimeter-wave on-chip spectrometer based on LEKID}

\maketitle

\section{Introduction}
In the framework of millimeter-wave astronomy, on-chip spectrometers aim to achieve intermediate spectral resolution, i.e. $R = \lambda / \Delta\lambda = 100 - 1000$, preserving the high sensitivity typical of continuum detectors. The key advantage compared to the existing Fourier transform spectrometers \citep{concerto2020, concerto2021} is the reduction of the optical background per channel, with a gain in the achievable photon-limited noise that can theoretically approach $1/\sqrt{R}$. On top of that, the optical design is enormously simplified. The main drawback on the other hand is the need of one readout channel per spectral band. With the currently available technology it is thus impossible to achieve large instantaneous field-of-views. 

On chip spectrometers based on Kinetic Inductance Detectors (KID) have already been proposed (Micro-Spec, \cite{microspec}, SPT-SLIM,\cite{SPT_SLIM},WSPEC, \cite{WSPEC},) and fabricated and tested ( SuperSpec, \cite{SuperSpec}, DESHIMA, \cite{DESHIMA}). The Micro-Spec spectrometer was designed for a resolution around 512 in the 420-540~GHz band. The SuperSpec device demonstrated a resolution of $\sim$275 in the 255-278 GHz band. The DESHIMA spectrometer on the ASTE 10-meters telescope achieved a bandwidth of 45~GHz centered around 350 GHz with a resolution of 380. 

In this work, we present the design, fabrication and first tests of an on-chip spectrometer based on Lumped Kinetic Inductance Detectors (LEKID). Our OMKID spectrometer targets the relevant 85-110~GHz atmospheric window. The concept is similar to that of the spectrometers mentioned above. The main differences are: a) the frequency band; b) the horn-coupling allowing a better control of the beam; c) the mono-crystalline microstrip configuration that simplifies the processing, ensures low RF losses and protects the device from spurious, i.e. unfiltered, radiation. 
The OMKID fabrication process has been designed to be compatible with large composite (imaging and spectroscopy) focal planes. That will allow to preserve the large mapping speed and at the same time spectroscopically image smaller/selected portions of the overall field-of-view.

\section{Design and Fabrication}

\begin{figure*}
\begin{center}
\resizebox{14cm}{!}{\includegraphics{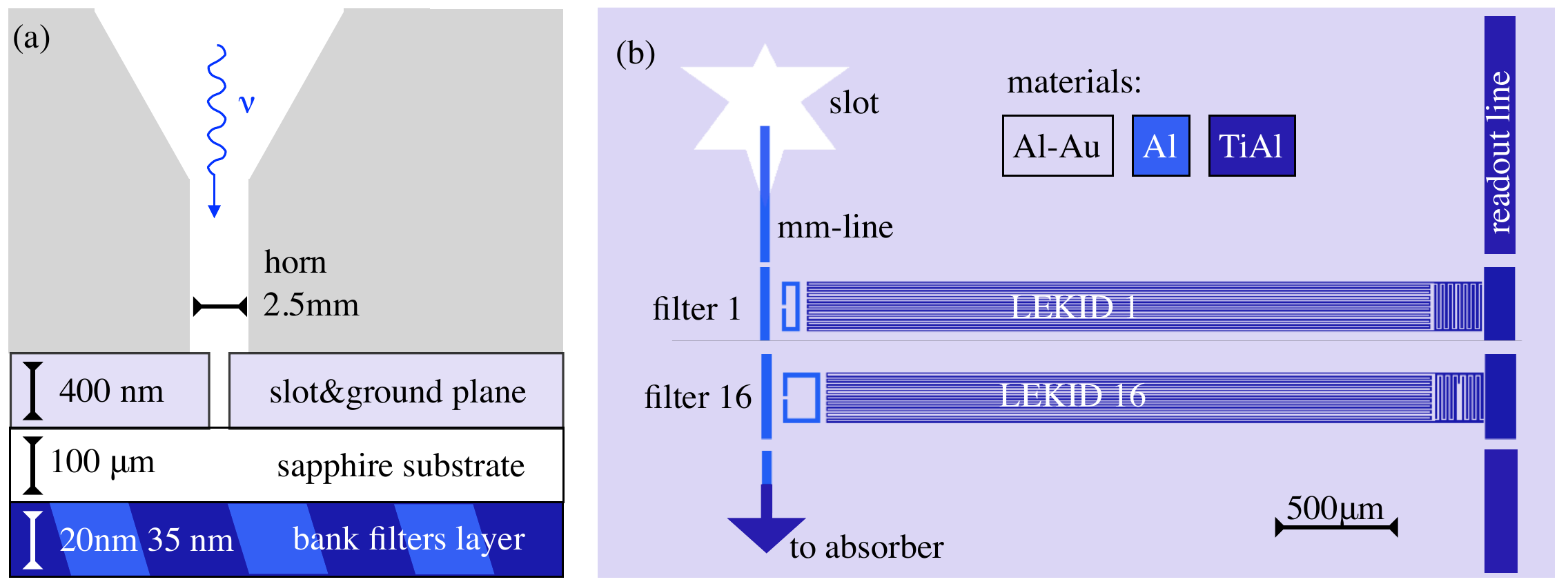}}
\caption{\textbf{OMKID spectrometer schematic} The spectrometer targets the 85-110~GHz range with 16 channels. (a) Side view. The radiation is collected through a horn-waveguide, a slot termination, a sapphire substrate and a mm-wave transmission line. (b) Top view. The c-shape filters pick up the signal from the mm-wave line. The filtered signal is absorbed by the facing LEKID made of TiAl (with a spectroscopy gap of 70~GHz). The resonance frequencies of the LEKID are monitored with the GHz readout line.}
\label{fig_design}
\end{center}
\end{figure*}

Our OMKID spectrometer has been designed and fabricated on a two inches 100~$\mu m$-thick monocrystalline C-plane sapphire substrate. A schematic view is presented in figure~\ref{fig_design}.
The incoming signal, centered around 90~GHz, is collected by a smooth conical horn antenna with a flare angle of 30 degrees. The 12~mm long horn ends up in a portion of cylindrical waveguide with a diameter of 2.5~mm. The cutoff frequency is around 70~GHz, and the antenna is expected to be single-mode ($TE_{11}$) until the onset of the $TM_{01}$ mode above 91~GHz. The waveguide itself is terminated by a star-shaped slot etched in the ground plane that is deposited on the back side of the sapphire substrate. The slot is fed by a mm-wave microstrip line placed on the front side. The transmission line guides the signal up to the filters and ends up in an absorber. The absorber aims to suppress possible standing waves. The co-planar superconducting filters are coupled to the LEKID that dissipate the mm-wave signal via generation of quasi-particles. The resonance frequencies of the LEKID are affected by the change of their kinetic inductance. A single GHz (microstrip) readout line is frequency-multiplexing the sixteen channels on a single electronics IN/OUT pair of coaxial cables. The readout rate for the OMKID tests is fixed at 46~Hz.

Our star-shape slot is only one among the possible waveguide-microstrip transitions. Similar fractal designs had already been proposed to operate at lower frequencies and as free-space coupled antennas \citep{antenna}. From our simulations, even after matching the microstrip feed, we estimate an optimal operation in the range 85-95~GHz. The waveguide-microstrip transition is thus the bottleneck of our design in terms of overall bandwidth.

The superconducting filters are properly-shaped half wave planar resonators. The length of each of the sixteen filters is varied to cover the 85-110~GHz range. The coupling quality factor of each filter with the mm-wave microstrip is adjusted to target a bandwidth of about 0.2~GHz. The notable c-shape of the filters aims at maximizing the current next to the LEKID, and optimising the dissipation efficiency in the detector. 

Each LEKID consists of a meandered inductor and an interdigitated capacitor. The resonant frequency is tuned by adjusting the length of the meander (L) and one of the fingers of the capacitor (C). The resonance frequencies range between 1.5~GHz and 1.8~GHz. The internal quality factor $Q_i$ (material) and the coupling quality factor $Q_c$ (design) are both in the $10^5$ range.
The LEKID, as in our standard imaging devices \citep{nika2}, are coupled to a 50~$\Omega$ micro-strip readout-line.

Three types of superconducting materials have been used to fabricate the OMKID: pure aluminum (20nm), titanium-aluminum bi-layer (10nm/25nm) and aluminum cover with gold (200nm/200nm). 
The pure aluminum superconductor (Al) is a lossless conductor for frequencies smaller than its spectroscopy gap $2\Delta_{Al}/h\sim$110~GHz. It is used for the (lossless) mm-wave microstrip and the filters and patterned via standard UV lithography and lift-off. The titanium-aluminum bi-layer (TiAl) aims to dissipate the radiation for frequencies higher than its spectroscopic gap $2\Delta_{TiAl}/h\sim$70~GHz \citep{catalano2015}. It is used for the LEKID and the absorber. To ensure the repeatability of the LEKID coupling quality factor $Q_c$ against possible lithography misalignments, the TiAl bi-layer is also used for the GHz microstrip readout line. This layer is also patterned using standard UV lithography followed by diluted HF etching. The aluminum covered with Au is used for the ground plane on the back-side of the substrate. The Au layer is crucial to ensure a good thermalisation of the device (and thus a fast time constant).

\section{Experimental set-up}
The on-chip spectrometer is mounted in a custom optical dilution refrigerator (mm-wave camera) with a base temperature of about 80~mK. The camera is directly derived from the NIKA (N\'eel IRAM KID Arrays) instrument \citep{monfardini2011}. The cold optics is composed by three high-density polyethylene (HDPE) smooth lenses (L1,L2,L3). L1 is located at room temperature and coincides with the cryostat vacuum isolation window. L2 is mounted on the screen at a temperature of about 4~K, while the final L3 lens is installed at the coldest stage just in front of the OMKID horn. Several custom low-pass filters installed on the radiations screens at 100~K, 50~K, 4~K, 1~K, 0.1~K ensure an optimal rejection of the out-of-band radiation. The background on the device is set by the diameter of the cold (0.1~K) pupil lying between L2 and L3.

For characterisation, two radiation sources have been used: i) a room-temperature black body (bb-300K) or a mirror allowing to "look inside" the cryostat, for an equivalent background temperature approaching 0~K (bb-0K); ii) a commercial mm-wave source coupled to a pyramidal emission horn (mm). The mm-wave source is polarised and can be set in the range 75-110~GHz with Hz-like frequency precision. It is, however, not calibrated in intensity, requiring a relative comparison with a well known and modeled black-body. 

The signal per channel of the spectrometer is defined as the frequency shift of the corresponding LEKID. In a previous publication, we have demonstrated the linear proportionality between the absorbed power and the frequency shift \citep{swenson2010}. The frequency shift of the LEKID is measured either with a vector network analyzer (VNA) or using a dedicated multiplexing electronics \citep{bourrion2013}, synchronized with the mm-source.

\section{Results and discussion}

\begin{figure}
\begin{center}
\resizebox{\linewidth}{!}{\includegraphics{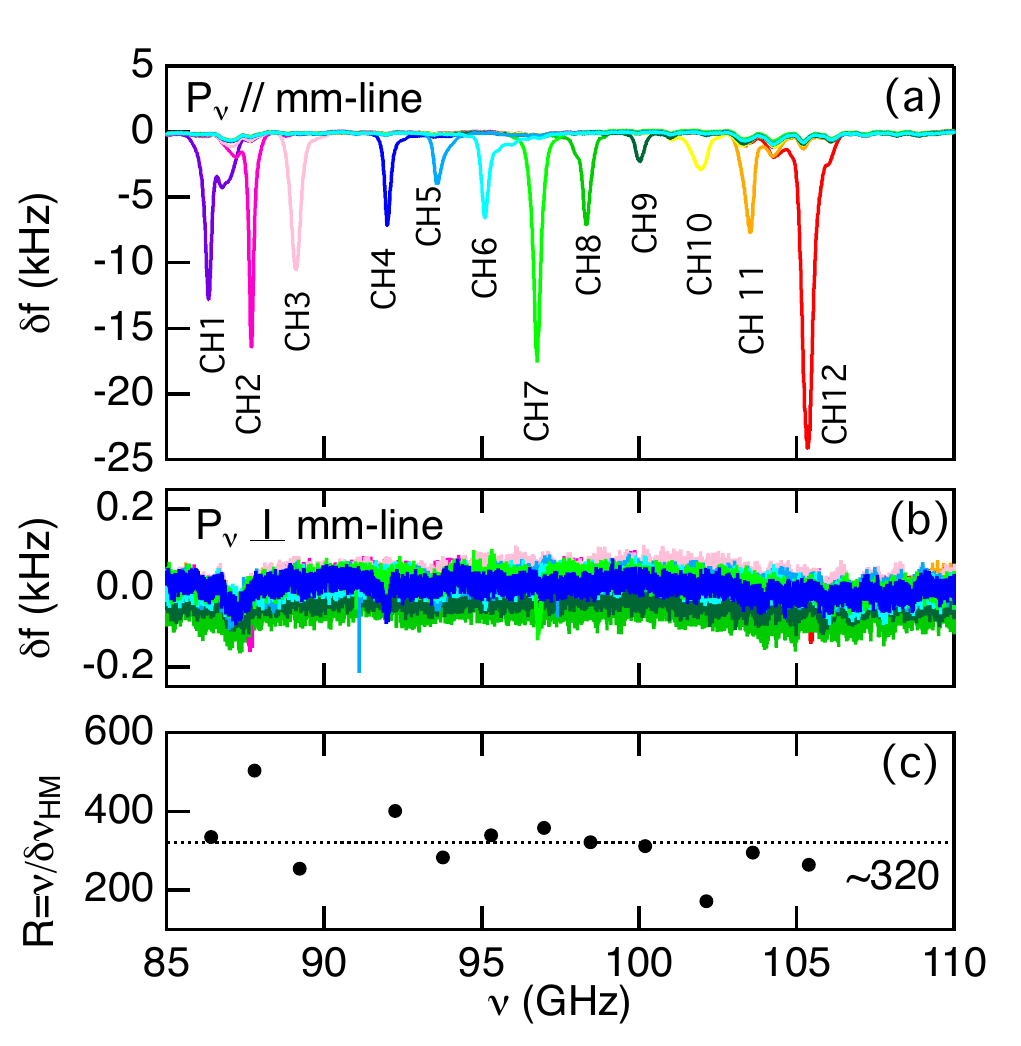}}
\caption{\textbf{OMKID spectrometer channels responses.} Frequency shift of the LEKID per channel as a function of the radiation frequency:  (a) for a polarisation of the mm-source parallel to the mm-wave line (b) for the perpendicular polarisation. (c) Spectral resolution per channel.}
\label{fig_channels}
\end{center}
\end{figure}
Figure~\ref{fig_channels} displays the OMKID spectrometer channel responses. Twelve out of the sixteen designed channels are functional. As expected from simulations, the device is polarisation selective. Down to the noise level, no signal is detected for a polarisation perpendicular to the mm-wave line. 
The upper limit on the cross-polarization is only 0.8\% for the best channel (\#12).



\begin{figure}
\begin{center}
\resizebox{\linewidth}{!}{\includegraphics{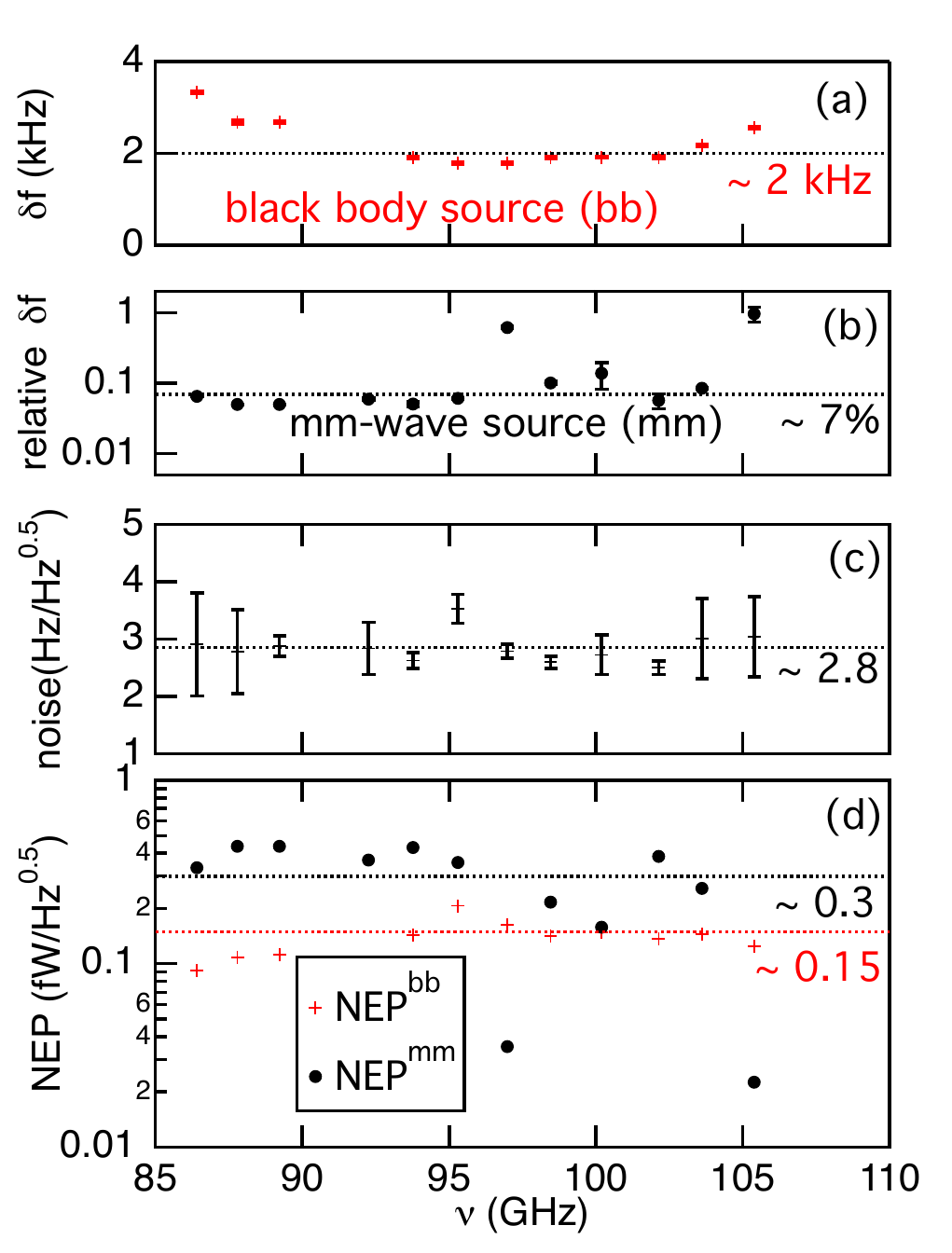}}
\caption{\textbf{Noise Equivalent Power Estimation.} As a function of the channel frequency: (a) LEKID frequency shift for illuminations with black body sources at 300~K and 0~K; (b) LEKID relative frequency shift for an illumination with the mm-wave source. Normalisation with respect to state-of-the art NIKA pixels (see text for more details). (c) LEKID noise spectral density at 10~Hz. (d) Noise Equivalent Power (NEP) estimated with the black body source illumination (bb) or with the mm-wave source (mm).}
\label{fig_NEP}
\end{center}
\end{figure}

Figure~\ref{fig_NEP} shows the measured optical Noise Equivalent Power (NEP) per channel and the elements used to evaluate it. The NEP corresponds to the power producing a signal-to-noise ratio of one in one Hz output bandwidth. It is calculated as: 
\begin{eqnarray}
\label{eq_NEP}
\textrm{NEP}=\frac{\Delta W_{\textrm{opt}}N}{\delta f}
\end{eqnarray}
where $\delta f $ is the frequency shift of the LEKID generated by the change of the optical load $\Delta W_{\textrm{opt}}$ (the power to be detected) and $N$ is the noise spectral density. 
As explained previously, the mm-wave source intensity is not calibrated. On top of that, and according to our previous experiences, we concluded that two independent methods are preferable for cross-checking the optical sensitivity estimations.

In the \textbf{first approach} we measure the frequency shift per channel for the 0K-300K temperature change of the black body source (fig.~\ref{fig_NEP}, panel a), and estimate the corresponding change of the optical load using a three-dimensional ray-tracing software, where the inputs are the spectral luminescence of the black-body source and the geometry of the cryostat (the apertures, the distances, the lenses curvatures and materials). The effective collecting surface of the horn is around 30~mm$^2$. The optical load varies from 150~fW for the 80-81~GHz band to 210~fW for the 100-101~GHz band for both  polarisations. In order to be conservative we use $\Delta W_{\textrm{opt}}\sim$ 105~fW (for one polarization). This value, together with the measure of the noise spectral density at 10~Hz (fig.~\ref{fig_NEP}, panel c), leads to $\textrm{NEP}^{bb}\sim0.15$~fW/Hz$^{0.5}$ (panel d).
With this approach, the NEP value is underestimated as the frequency shift is not only due to the channel response but also to the integrated background response. From lorentzian plus constant background fit of the OMKID spectrometer channel responses (figure~\ref{fig_channels}, panel a) we estimate the integrated background response to contribute in average to almost half the total response. For the best channel (\#12), the integrated background response contributes to about 20\% of the total response. 

In the \textbf{second approach} we employ the mm-wave source to illuminate the OMKID, and compare its response with that of a well-characterized NIKA-like imaging array made of the same material \citep{catalano2020}, which we mount and test in the same optical configuration and using the same source. For each channel, we evaluate the relative frequency shift as the ratio of the signal of the OMKID to that of the NIKA array, both measured at the peak frequency of the channel considered (fig.~\ref{fig_NEP}, panel b). A geometrical correction factor has been applied, equal to the ratio of the effective collecting surface of the horn ($\approx~30~mm^2$) to the surface of the NIKA pixels ($5.3~mm^2$).
The OMKID and the NIKA pixels exhibit similar noise and resonance frequencies values. Thus, assuming 100\% quantum efficiency for the NIKA pixels, the relative frequency shift corresponds to the OMKID overall optical efficiency.
The optical NEP of the NIKA array is, on average, $2.2\times10^{-17}$W/Hz$^{0.5}$. Using this value and the relative frequency shift we then get $\textrm{NEP}^{mm}\sim0.3$~fW/Hz$^{0.5}$ (fig.~\ref{fig_NEP}, panel d). 

The two methods give consistent results, confirming also that the sensitivity is currently limited, as is the case for competing devices of this kind, by the relatively poor overall quantum efficiency.

\section{Conclusions}
We have developed an on-chip spectrometer based on LEKID and able to cover the astronomically relevant band 85-110~GHz. Our device is characterised by a relatively simple design and fabrication process, compatible with hybridisation on large field-of-view LEKID imaging arrays. The first prototype showed good performance in terms of spectral resolution ($R \approx 300$, as designed) and polarisation selection. The optical sensitivity figures are promising with an overall quantum efficiency in the range 5-10\% that is in line with existing devices performance. A significant improvement is possible as the overall efficiency is, according to simulations, limited to 20\% by the efficiency of the prototype waveguide-microstrip transition.

\section*{Acknowledgments}
We acknowledge the specific contribution of the engineer G.~Garde to the device holder and  the overall support of the Cryogenics, Electronics and Nanofab groups at Institut N\'eel and LPSC. 
The fabrication of the device described in this paper was conducted at the PTA Grenoble micro-fabrication facility. 
This work has been partially supported by the French National Research Agency through Grant No. ANR-16-CE30-0019 ELODIS2, the LabEx FOCUS through Grant No. ANR-11-LABX-0013 and the EU\textsc{\char13}s Horizon 2020 research and innovation program under Grant Agreement No. 800923 (SUPERTED).

\bibliographystyle{aa}
\bibliography{OMKID_v2_arxiv}

\end{document}